\documentclass[aps,onecolumn,showpacs,nofootinbib,showkeys]{revtex4-2}
\usepackage[margin=2.5 cm]{geometry}

\RequirePackage[T1]{fontenc}

\usepackage{epsfig,graphicx,amsmath,amssymb,bm}
\RequirePackage{mathptmx}      
\usepackage{mathrsfs}
\usepackage{amssymb}
\RequirePackage{color}
\usepackage{changes}
\usepackage{slashed}
\usepackage{multirow}
\usepackage{scalerel}
\usepackage{tikz-feynman}
\usepackage{textcomp}
\usepackage{subcaption}
\usepackage[english]{babel}
\usepackage{float}
\RequirePackage{hyperref}
\hypersetup{
    linktocpage,
    colorlinks,
    citecolor=blue,
    filecolor=black,
    linkcolor=blue,
    urlcolor=blue,
}
\usepackage{nccmath}

\begin{document}

\title{On the decay widths of radially excited scalar meson $K^*_0(1430)$ in view of new experimental data
}


\author{Mikhail K. Volkov$^{1}$}\email{volkov@theor.jinr.ru}
\author{Kanat Nurlan$^{1,2,3}$}\email{nurlan@theor.jinr.ru}
\author{Aleksey A. Pivovarov$^{1}$}\email{tex_k@mail.ru}

\affiliation{$^1$ Bogoliubov Laboratory of Theoretical Physics, JINR, 
                 141980 Dubna, Moscow region, Russia \\
                $^2$ The Institute of Nuclear Physics, Almaty, 050032, Kazakhstan\\
                $^3$ L. N. Gumilyov Eurasian National University, Astana, 010008, Kazakhstan}   


\begin{abstract}
Decays of scalar mesons $K^*_0(800) \to K\pi$ and $K^*_0(1430) \to K\pi, K \eta, K \eta', K_1 \pi$ are described in the extended $U(3 ) \times U(3)$ Nambu -- Jona-Lasinio chiral quark model. The obtained results are in satisfactory agreement with the new experimental data obtained by the BaBar collaboration, which markedly differ from the existing values in the PDG.


\end{abstract}


\maketitle


\section{\label{Intro}Introduction}
The description of scalar mesons in the ground and excited states is of great interest. In view of recent experimental data the study of the strange scalar meson $K^*_0$ in the ground and first radially excited states is especially interesting.

The scalar meson $K^*_0 (800)$ decays into a $K\pi$ pair with 100$\%$ probability \cite{ParticleDataGroup:2022pth}. At the same time, for the decay of the radially excited $K^*_0(1430)$ meson, the main channels are $K\pi$, $K\eta$ and $K\eta'$ \cite{ParticleDataGroup:2022pth}. The decay $K^*_0(1430) \to K\pi$ was first measured with sufficient accuracy in the study of the reaction $Kp \to K\pi n$ \cite{Aston:1987ir}. In a recent paper by the BaBar collaboration the decays of the charmed pseudoscalar meson $\eta_c$ were measured, where a detailed analysis of the data showed the decisive role of channels with scalar mesons \cite{BaBar:2021fkz}. This made it possible for the first time to measure the ratio of the branching fractions for the ${B(K^*_0(1430) \to \eta' K)}/{B(K^*_0(1430) \to \pi K)}$ decays and to estimate the coupling constants $g_{\eta' K}$ , $g_{\pi K}$. In this case, the $K^*_0(1430) \to K\pi$ decay widths obtained using the latter constant turns out to be noticeably smaller than the existing data in PDG \cite{ParticleDataGroup:2022pth}. Such discrepancy makes the issue of a theoretical estimation of the $K^*_0(1430)$ meson decay widths of topical.

In the present paper, we calculate the decay widths of radially excited scalar meson $K^*_0(1430) \to K\pi$, $K^*_0(1430) \to K\eta$, $K^*_0(1430) \to K\eta'$ and $K^*_0(1430) \to K_1 \pi$ within the extended $U(3) \times U(3)$ NJL chiral quark model \cite{Volkov:1996br, Volkov:1996fk, Volkov:1999yi, Volkov:2005kw, Volkov:2017arr, Volkov:2022jfr}. Regarding the last decay, in \cite{Roca:2021bxk} it was shown that the existing PDG data on the width of $K_1(1270) \to K^*_0 (1430) \pi$ imply $K^*_0 (1430) \to K_1( 1270) \pi$. We agree with this statement and give an independent estimation for the decay width of $K^*_0 (1430) \to K_1(1270) \pi$. However, the width $\Gamma(K^*_0 (1430) \to K_1(1270) \pi) = 40$ keV obtained by us turns out to be noticeably smaller than the width $\Gamma(K^*_0 (1430) \to K_1(1270) \pi)=2075 (+4100, -1100)$ MeV obtained in \cite{Roca:2021bxk}. Our result confirms the fact that the $K^*_0 (1430)$ meson width is practically exhausted by the $K\pi$, $K\eta$ and $K \eta'$ channels.

For the calculation of the $K^*_0(1430) \to [K\pi, K\eta, K\eta', K_1 \pi]$ decays, we use the extended NJL model \cite{Volkov:1996br, Volkov:1996fk, Volkov:1999yi, Volkov:2005kw, Volkov:2022jfr}. The NJL chiral quark model successfully describes interactions of four meson nonets of scalar, pseudoscalar, vector, and axial-vector types in the ground and first radially excited states using a limited number of fixed parameters. In the model, the effective chiral quark-meson Lagrangians are obtained, and the processes of meson production in $\tau$ decays and $e^+e^-$ annihilations, as well as numerous decays of radially excited mesons, are successfully described.

In our version of the $U(3) \times U(3)$ chiral NJL model, the value of the cutoff parameter is $\Lambda_4 =1250$ MeV \cite{Volkov:2022jfr}. This makes it possible to include in the model, in addition to the 4-meson nonets in the ground states, also their first radially excited states, and in this case, one can hope to obtain satisfactory results at a qualitative level in the framework of partial chiral symmetry conservation. Such an attempt was made in the works \cite{ Volkov:1996fk, Volkov:1999yi, Volkov:1999xf, Volkov:2005kw}. Radially excited states were described by introducing into the model the simplest form factor quadratic in the transverse momentum of quarks. In this case, the mixing of mesons in the radially excited state with the ground states was also taken into account, which leads to the appearance of off-diagonal terms in the free Lagrangian. These terms are diagonalized using mixing angles or a matrix in the case of $\eta, \eta', \eta(1295)$ and $\eta(1475)$ mesons. It is important to note that the slope parameter $d$ introduced in the extended model is fixed without using experimental data, based on the requirement that the quark condensate remains unchanged after radially excited states are taken into account. In this case, the values of quark masses and the ultraviolet cutoff parameter do not change.    

It is interesting to note that within the extended NJL model, the mass spectrum of the ground and excited scalar nonets was satisfactorily described (19 states of scalar mesons, taking into account mixings of five states: 4 scalar and one glueball) \cite{Volkov:1999yi, Volkov:2001ct}.

\section{Effective quark-meson Lagrangians of the NJL model}
The quark-meson Lagrangian for the strong interaction of scalar, pseudoscalar and axial vector mesons necessary for describing the processes considered here in the NJL model takes the form \cite{Volkov:1999xf, Volkov:1999yi, Volkov:2005kw, Volkov:2022jfr}
\begin{eqnarray}
{\cal L}_{int} = \bar{q} \biggl[ 
i \gamma_5 A_\pi \sum_{i = \pm, 0} \lambda^\pi_i \pi^i + i \gamma_5 A_K \sum_{i = \pm, 0} \lambda^K_i K^i
+ \frac{1}{2} \gamma_\mu \gamma_5 A_{K_1}  \sum_{i = \pm, 0} \lambda^K_i K^i_{1\mu} \\ \nonumber
+ \sum_{i = \pm} \lambda_{K}(A_{K^*_0} K^{*i}_0 + B_{K^*_0} \hat{K}^{*i}_0) + i\gamma^{5} \sum_{i = u, s} \lambda_{i} \left[A^{i}_{\eta}\eta + A^{i}_{\eta'}\eta' + A^{i}_{\hat{\eta}}\hat{\eta} + A^{i}_{\hat{\eta}'}\hat{\eta}'\right]
\biggl]q,
\end{eqnarray}
where $q$ and $\bar{q}$ are u, d and s quark fields with constituent quark masses $m_{u} \approx m_{d} = 270$~MeV, $m_{s} = 420$~MeV; excited mesonic states of mesons are marked with a hat and $\lambda$ are linear combinations of the Gell-Mann matrices \cite{Volkov:2022jfr},
\begin{eqnarray}
\label{verteces1}
	A_{M} = A^0_{M} \left[g_{M}\sin\theta^+_{M} +
	g'_{M}f_{M}(k_{\perp}^{2})\sin\theta^-_{M}\right], \nonumber\\
	B_{M} = - A^0_{M} \left[g_{M}\cos\theta^+_{M} +
	g'_{M}f_{M}(k_{\perp}^{2})\cos\theta^-_{M}\right],
\end{eqnarray}
where $A^0_{M} = 1/{\sin(2\theta_{M}^{0})}$ and $\theta^{\pm}_{M} = \theta_M \pm \theta^0_M$. The subscript M indicates the corresponding meson; $\theta_{\pi} = 59.48^{\circ}, \theta_{\pi}^{0} = 59.12^{\circ}, \theta_{K} = 58.11^{\circ}$, $\theta_{K}^{0} = 55.52^{\circ}$, $\theta_{K_1} = 85.97^{\circ}$, $\theta_{K_1}^{0} = 59.56^{\circ}$, $\theta_{K^*_0} = 74.0^{\circ}$ and $\theta_{K^*_0}^{0} = 60.0^{\circ}$ are the mixing angles \cite{Volkov:2022jfr}. The mixing angles for the K and $\pi$ mesons $\theta \approx \theta_0$; so for the ground states of these mesons one can use $A_\pi = g_\pi$ and $A_K=g_K$. 

For the $\eta$ mesons, the factor $A$ takes a slightly different form. This is due to the fact that 
in the case of the $\eta'$ meson four states are mixed 
        \begin{eqnarray}
            A^{u}_{M} & = & g_{\eta^{u}} a^{u}_{1M} + g'_{\eta^{u}} a^{u}_{2M} f_{uu}(k_{\perp}^{2}), \nonumber\\
            A^{s}_{M} & = & g_{\eta^{s}} a^{s}_{1M} + g'_{\eta^{s}} a^{s}_{2M} f_{ss}(k_{\perp}^{2}).
        \end{eqnarray}
        
Here $f\left(k_{\perp}^{2}\right) = \left(1 + d k_{\perp}^{2}\right)\Theta(\Lambda^{2} - k_{\perp}^2)$ is 
the form-factor describing the first radially excited meson states. The slope parameters, 
$d_{uu} = -1.784 \times 10^{-6} \textrm{MeV}^{-2}$ and $d_{ss} = -1.737 \times 10^{-6} \textrm{MeV}^{-2}$, 
are unambiguously fixed from the condition of constancy of the quark condensate after the inclusion 
of radially excited states and depends only on the quark composition of the corresponding meson \cite{Volkov:2022jfr}.

The values of the mixing ($A$) parameters are shown in Table \ref{tab_eta}. The $\eta'$ meson corresponds to the physical state $\eta'(958)$ and the $\hat{\eta}$, $\hat{\eta}'$ mesons correspond to the first radial excitation mesons $\eta$ and $\eta'$.
    
\begin{table}[h!]
\begin{center}
\begin{tabular}{ccccc}
\hline
   & $\eta$ & $\hat{\eta}$ & $\eta'$ & $\hat{\eta}'$ \\
\hline
$a^{u}_{1}$		& 0.71			& 0.62            &-0.32             & 0.56    \\
$a^{u}_{2}$		& 0.11			& -0.87           & -0.48            & -0.54   \\
$a^{s}_{1}$               & 0.62                        & 0.19            & 0.56             & -0.67 \\
$a^{s}_{2}$               & 0.06                       & -0.66           & 0.3               & 0.82 \\
\hline
\end{tabular}
\end{center}
\caption{Mixing parameters of $\eta$ mesons \cite{Volkov:1999yi, Volkov:2022jfr}.}
\label{tab_eta}
\end{table}   
    
The quark-meson coupling constants have the form
\begin{eqnarray}
	\label{Couplings}
 g_{\pi} = g_{\eta^{u}}=\left(\frac{4}{Z_{\pi}}I_{20}\right)^{-1/2}, 
\, g'_{\pi}=g'_{\eta^{u}} =  \left(4 I_{20}^{f^{2}}\right)^{-1/2}, 
\, g_{K} =\left(\frac{4}{Z_K}I_{11}\right)^{-1/2}, \nonumber\\
\, g'_{K} =\left(4I^{f^2}_{11}\right)^{-1/2},  
g_{K_1} =\left(\frac{2}{3}I_{11}\right)^{-1/2}, 
\, g'_{K_1} =\left(\frac{2}{3}I_{11}^{f^{2}}\right)^{-1/2}, \nonumber\\
 g_{K^*_0} =\left(4I_{11}\right)^{-1/2}, 
\, g'_{K^*_0} =\left(4I^{f^2}_{11}\right)^{-1/2}, 
\end{eqnarray}
here $Z_{\pi}$ and $Z_{\eta^{s}}$ are additional renormalization constants appearing 
in the pseudoscalar and axial-vector transitions \cite{Volkov:2005kw, Volkov:2022jfr}.

Integrals appearing in the quark loops are
\begin{eqnarray}
	I_{n_{1}n_{2}}^{f^{m}} =
	-i\frac{N_{c}}{(2\pi)^{4}}\int\frac{f^{m}(k^2_{\perp})}{(m_{u}^{2} - k^2)^{n_{1}}(m_{s}^{2} - k^2)^{n_{2}}}\Theta(\Lambda_{3}^{2} - k^2_{\perp})
	\mathrm{d}^{4}k,
\end{eqnarray}
where $\Lambda_3=1030$ MeV is is the three-dimensional cutoff parameter, the value of four-dimensional cutoff parameter is $\Lambda_4=1250$ MeV \cite{Volkov:2005kw}. 

When describing decays involving $K_1$ axial vector mesons, we take into account the mixing effect of the $K_{1A}$ and $K_{1B}$ states \cite{Volkov:1986zb, Suzuki:1993yc}. The mixing of the axial vector mesons $K_{1A}$ and $K_{1B}$ leads to physical states $K_1 (1270)$ and $K_1 (1400)$ \cite{ParticleDataGroup:2022pth}. This mixing is described as follows:
	\begin{eqnarray}
	\label{K1AK1B}
		K_{1A} & = & K_{1}(1270)\sin{\alpha} + K_{1}(1400)\cos{\alpha}, \nonumber\\
		K_{1B} & = & K_{1}(1270)\cos{\alpha} - K_{1}(1400)\sin{\alpha},
	\end{eqnarray}
where $\alpha=57^{\circ}$	\cite{Volkov:2022jfr}. This effect was also considered in the works \cite{Volkov:1984gqw, Li:1996md, Li:2005eq, Geng:2006yb, Cheng:2013cwa}. 

\section{Amplitudes and decay widths} 
\begin{figure*}[t]
   \centering
    \begin{tikzpicture}
     \begin{feynman}
      \vertex (a) {\(K^*_0 \)};
      \vertex [dot, right=1.6cm of a] (b) {};
      \vertex [dot, above right=2cm of b] (c) {};
      \vertex [dot, below right=2cm of b] (d) {};
      \vertex [right=1.6cm of c] (g) {\(K\)};
      \vertex [right=1.6cm of d] (f) {\(\pi \)};
      \diagram* {
        (a) -- [] (b),
        (b) -- [fermion] (c),
        (c) -- [fermion] (d),
        (d) -- [fermion] (b),  
        (c) -- [] (g),         
        (d) -- [] (f),
      };
     \end{feynman}
    \end{tikzpicture}
   \caption{Triangle quark diagram for the decay of scalar meson $K^*_0 \to K \pi$}
 \label{diagram1}
\end{figure*}
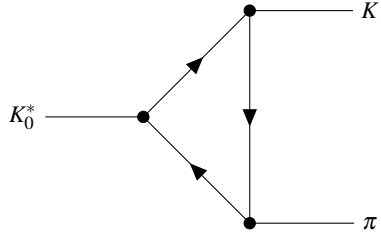

We start with considering the decay $K^*_0(800) \to K \pi$. This process is described by the quark diagram given in Figure \ref{diagram1}. Quark loops are calculated using the methods developed in the NJL model and successfully tested on other physical processes \cite{Volkov:2005kw, Volkov:2022jfr}. The loop integrals are expanded in terms of the external fields momentums, and only the logarithmic divergent parts are preserved. Accounting for such terms makes it possible to preserve the chiral symmetry in the model \cite{Volkov:1986zb}. Model calculations lead to the following formula for the decay width $K^*_0(800) \to K \pi$:
\begin{eqnarray}
\label{k0800}
\Gamma(K^{*-}_0 \to K^-\pi^0) =
\frac{1}{2J_{K^*_0}+1} \frac{\left( 8m_s I^{K^*_0 K \pi}_{11}\right)^2}{2M_{K^*_0}} \frac{\sqrt{E^2_K - M^2_K}}{4\pi M_{K^*_0}},
\end{eqnarray}
where 
\begin{eqnarray}
J_{K^*_0}=0, \,E^2_K = \frac{M^2_{K^*_0} + M^2_K - M^2_{\pi^0}}{2},
\end{eqnarray}
where the meson masses are taken from PDG \cite{ParticleDataGroup:2022pth}. 

The integrals
\begin{eqnarray}
\label{integral}
&& I_{n_1n_2}^{M M'...}(m_{u}, m_{s}) = -i\frac{N_{c}}{(2\pi)^{4}} 
 \int\frac{A(k_{\perp}^{2})...B(k_{\perp}^{2})...}{(m_{u}^{2} - k^2)^{n_1}(m_{s}^{2} - k^2)^{n_2}}
\Theta(\Lambda_{3}^{2} - \vec{k}^2) \mathrm{d}^{4}k,
\end{eqnarray}
are obtained from the quark triangular loops, $A(k_{\perp}^{2})$ and $B(k_{\perp}^{2})$ are the coefficients
for different mesons defined in (\ref{verteces1}).

The decay amplitude of $K^{*-}_0 \to K^0 \pi^-$ has a similar structure with an additional factor $\sqrt{2}$. As a result, for the width we obtain $\Gamma(K^{*-}_0 \to K^-\pi^0 +K^0 \pi^-) = 430$ MeV. Calculations in the standard NJL model lead to a close result $\Gamma(K^{*-}_0 \to K^-\pi^0 +K^0 \pi^-) = 450$ MeV. The experimental value for the width of this decay is $\Gamma(K^{*}_0 \to K\pi)_{exp} = 468 \pm 30$ MeV \cite{ParticleDataGroup:2022pth}.

To calculate the decay of a radially excited meson $K^*_0(1430) \to K \pi$ in the obtained amplitude (\ref{k0800}), we replace the vertex $K^*_0(800) \to K^*_0(1430) $. As a result, for the width and decay constant $g_{K^{*'}_0 \to K\pi}$ in the extended NJL model, we obtain
\begin{eqnarray}
\Gamma(K^*_0(1430) \to K\pi)_{NJL} = 18.543 \, \text{MeV}, \quad g^2_{K\pi} = 0.515 \, \text{GeV}^2.
\end{eqnarray}

The model predictions for this decay can be compared with the data of the BaBar collaboration derived from the analysis of the processes $\gamma \gamma \to \eta_c \to (\pi, \eta')KK$ \cite{BaBar:2021fkz}
\begin{eqnarray}
g^2_{K\pi} = 0.458 \pm 0.032_{\text{stat}} \pm 0.044_{\text{sys}} \, \text{GeV}^2.
\end{eqnarray}

As we can see, the results of the calculation in the NJL model for the value of the constant agree satisfactorily with the new experimental data. The decay width obtained using the experimental value of the constant $g_{K\pi}$ is equal to $\Gamma(K^*_0(1430) \to K\pi) = 16.46 \pm 1.15$ MeV at the mass $M_{K^*_0(1430)} = 1425 \pm 50$ MeV. It turns out to be less than the width $\Gamma(K^*_0(1430) \to K\pi) = 251.10 \pm $27.0 MeV \cite{Aston:1987ir}, which is given in PDG.  

Our calculations show that the decay width of the radially excited meson $K^*_0(1430) \to K \pi$ is noticeably smaller than the width in the ground state $K^*_0(800) \to K \pi$. Note that a similar situation was observed in the NJL model when describing the decays $\rho \to 2\pi$ and $\rho' \to 2\pi$ \cite{Volkov:1999yi}.

Next, we consider the decays of $K^*_0(1430)$ with the production of meson pairs $K \eta$ and $K \eta'$. Here it is necessary to take into account both the $u$,$d$ and $s$ quark parts of these mesons. As a result, we obtain the following amplitudes
\begin{eqnarray}
\mathcal{M}(K^*_0(1430) \to K\eta)= 8m_s I^{\hat{K}^*_0 K \eta_u} - 8\sqrt{2}m_u I^{\hat{K}^*_0 K \eta_s}, \\
\mathcal{M}(K^*_0(1430) \to K\eta')= 8m_s I^{\hat{K}^*_0 K \eta'_u} - 8\sqrt{2}m_u I^{\hat{K}^*_0 K \eta'_s}.
\end{eqnarray}

Numerical estimates lead to the following values for the width and decay constants in the NJL model
\begin{eqnarray}
\Gamma(K^*_0(1430) \to K\eta)_{NJL} = 0.291 \text{MeV}, \nonumber\\
g^2_{K\eta} = 0.030 \, \text{GeV}^2, \quad g^2_{K\eta'} = 0.671 \, \text{GeV}^2. 
\end{eqnarray}

In the work, \cite{BaBar:2021fkz} the BaBar collaboration presented
\begin{eqnarray}
\frac{g^2_{K\eta'}}{g^2_{K\pi}} = 1.50 \pm 0.24_{\text{stat}} \pm 0.24_{\text{sys}}.
\end{eqnarray}

For this ratio in the NJL model we get $g^2_{K\eta'}/g^2_{K\pi} = 1.30$. These results can be considered satisfactory within the experimental and model accuracies. The model precision is estimated as $\pm15\%$ based on the statistical analysis of previous numerous calculations and partial axial current conservation (PCAC) \cite{Volkov:2022jfr}.

Next, consider the decay $K_1(1270) \to K^*_0(1430) \pi$. This process is possible in the case of $M_{K_1}(1270) > M_{K^*_0}(1430)+M_\pi$, which is unlikely and can be due to the large width of the mesons $K_1(1270)$ and $K^* _0(1430)$ and uncertainties in mass definitions $M_{K^*_0(1430)} = 1425 \pm 50 $ MeV and $M_{K_1(1270)} = 1253 \pm 7$ MeV \cite{ParticleDataGroup:2022pth}. In \cite{Roca:2021bxk}, the decay $K^*_0 (1430) \to K_1 \pi$ is described, which is a reversal reaction of $K_1(1270) \to K^*_0(1430) \pi$, and a relatively large width $\Gamma=2075 (+4100, -1100)$ MeV is obtained. This width turns out to be wider than the meson width $\Gamma_{K^*_0(1430)}=270 \pm 80$ MeV and does not correspond to the fact that the $K^*_0(1430)$ meson predominantly decays into $K \pi $ and $K \eta$ \cite{ParticleDataGroup:2022pth}. However, the authors did not attempt to give an exact value for the width but showed a discrepancy of the process $K_1(1270) \to K^*_0(1430) \pi$ given in the PDG. This decay in the NJL model is described by the amplitude
\begin{eqnarray}
\mathcal{M}(K^*_0(1430) \to K_1 \pi) = 2 \sin{\alpha}I^{\hat{K}^*_0 K_1 \pi}_{11} (p_{K^*_0}+p_\pi)_\mu \epsilon_\mu(p_{K_1}), 
\end{eqnarray}
where $\epsilon_\mu (p_{K_1})$ is the polarization vector of the $K_1$ meson with the momentum $p_{K_1}$; $p_{K^*_0}$ and $p_\pi$ are the momenta of the pion and scalar meson $K^*_0(1430)$. Accordingly, in the model, we obtain the decay width
\begin{eqnarray}
\Gamma(K^*_0(1430) \to K_1 \pi)_{NJL} = 40 \, \text{keV}.
\end{eqnarray}

\section{Conclusions}
In this paper, we have described the decay channels of scalar mesons in the ground $K^*_0(800)$ and first radially excited state $K^*_0(1430)$. Our results for the decays of the excited meson $K^*_0(1430)$ are in satisfactory agreement with the recent experimental data of the BaBar collaboration \cite{BaBar:2021fkz} and at the same time deviate noticeably from the PDG data \cite{ParticleDataGroup:2022pth}.

At the present time, in describing scalar mesons, an important role is played by the assumption of the tetraquark structure of scalar mesons \cite{Achasov:1999wv}. This especially concerns isovector scalar mesons $a_0$, where it is impossible to correctly describe the mass $M_{a_0(980)} = 980 \pm 20$ MeV without the assumption of a tetraquark structure. At the same time, the masses of the isoscalar mesons $f_0(980)$, as well as strange mesons $K^*_0$, both in the ground and first radially excited states, are quite satisfactorily described based on the quark-antiquark structure \cite{Volkov:1999yi}. This allows us to assume in this paper that the strange mesons $K^*_0(800)$ and $K^*_0(1430)$ mainly have a quark-antiquark structure. The presence of quark-antiquark structures in other scalar mesons also follows from the existence of chiral symmetry. There are a number of works that evaluate the relative role of quark-antiquark and tertraquak states in determining the $a_0(980)$ structure \cite{tHooft:2008rus, Kim:2017yvd, Lee:2019bwi}

As regards the $K_1(1270) \to K^*_0(1430) \pi$ decay, we agree with the statement of the authors of \cite{Roca:2021bxk} that the existing PDG data on the partial width of the $K_1(1270) \to K^*_0(1430) \pi$ imply the process $K^*_0(1430) \to K_1(1270) \pi$. In addition, we give an independent estimate of the $K^*_0(1430) \to K_1(1270) \pi$ decay width, which turns out to be much smaller than the width obtained in \cite{Roca:2021bxk} and gives a small contribution to the total width of $\Gamma_{ K^*_0(1430)}$.

\subsection*{Acknowledgments}
We are grateful to prof. A.B. Arbuzov for his interest in our work and important remarks which 
improved the paper.


\end{document}